\RequirePackage{fix-cm}
\documentclass{svjour3}                     
\smartqed  
\usepackage{appendix}
\usepackage{amsmath}
\usepackage{graphicx}
\usepackage{lineno}
\usepackage{array}
\usepackage{longtable}
\usepackage{natbib}
\setcitestyle{aysep={}}
\nolinenumbers
\newcommand*\patchAmsMathEnvironmentForLineno[1]{%
\expandafter\let\csname old#1\expandafter\endcsname\csname #1\endcsname
\expandafter\let\csname oldend#1\expandafter\endcsname\csname end#1\endcsname
\renewenvironment{#1}%
{\linenomath\csname old#1\endcsname}%
{\csname oldend#1\endcsname\endlinenomath}}%
\newcommand*\patchBothAmsMathEnvironmentsForLineno[1]{%
\patchAmsMathEnvironmentForLineno{#1}%
\patchAmsMathEnvironmentForLineno{#1*}}%
\AtBeginDocument{%
\patchBothAmsMathEnvironmentsForLineno{equation}%
\patchBothAmsMathEnvironmentsForLineno{align}%
\patchBothAmsMathEnvironmentsForLineno{flalign}%
\patchBothAmsMathEnvironmentsForLineno{alignat}%
\patchBothAmsMathEnvironmentsForLineno{gather}%
\patchBothAmsMathEnvironmentsForLineno{multline}%
}

\usepackage{xcolor}
\usepackage{amssymb}
\usepackage{upgreek}

\newcommand{\dd}{\ensuremath{\mathrm{d}}}
\newcommand{\df}[2]{\ensuremath{\f{\dd #1}{\dd #2}}}
\newcommand{\pf}[2]{\ensuremath{\frac{\partial{#1}}{\partial{#2}}}}
\newcommand{\pfsq}[2]{\ensuremath{\frac{\partial^2{#1}}{\partial{#2}^2}}}
\newcommand{\f}[2]{\ensuremath{\frac{#1}{#2}}} 
 
\newcommand{\id}{\ensuremath{\hspace{0.6mm}\mathrm{d}}}

 \mathchardef\mhyphen="2D
\newcommand{\mb}[1]{\ensuremath{\mathbf{#1}}}

\definecolor{myred}{RGB}{163, 20, 47}
\definecolor{myblue}{RGB}{0, 115,190}
\definecolor{mypurp}{RGB}{127,47,143}
\definecolor{mygold}{RGB}{239,178,32}
\definecolor{mygreen}{RGB}{0,129,0}
\definecolor{myorang}{RGB}{218,84,25}

\begin{document}

\title{
Direct Numerical Simulation of 
the Moist Stably Stratified Surface Layer:
Turbulence and Fog Formation
}

\titlerunning{DNS of the Moist SBL: Turbulence and Fog Formation}        

\author{Michael MacDonald\text{*}        \and
        Marcin J. Kurowski \and
              Jo{\~{a}}o Teixeira
}


\institute{
\text{*}M. MacDonald \at
 Jet Propulsion Laboratory, California Institute of Technology, Pasadena, CA, USA 
 and
 Department of Mechanical Engineering, The University of Auckland, Auckland, New Zealand\\
              \email{michael.macdonald@auckland.ac.nz}           
\and
M. J. Kurowski  \and J. Teixeira \at
              Jet Propulsion Laboratory, California Institute of Technology, Pasadena, CA, USA \\
}

\date{Received: DD Month YEAR / Accepted: DD Month YEAR}

\maketitle

\fontfamily{ptm}\selectfont

\begin{abstract}
We investigate the effects of condensation and liquid water loading  on the stably stratified surface layer, with an eye towards understanding the influence of turbulent mixing on fog formation.
Direct numerical simulations (DNS) of dry and moist open channel flows are conducted, where in both a constant cooling rate is applied at the ground to mimic longwave radiative cooling.  
 Depending on the cooling rate, it can lead to either turbulent (weakly stable) or laminar (very stable) flows. 
Compared to the completely dry case, the condensation of liquid water  in the moist case enables slightly higher cooling rates to be achieved before leading to turbulence collapse.
In the very stable cases, runaway cooling  leads to the substantial condensation of liquid water close to the ground and fog (visibility less than 1 km) results over much of the domain. In the weakly stable cases, turbulent mixing narrowly yields visibilities of 1 km close to the ground over a similar time period. However, despite the idealized nature of the system, the present results suggest that turbulence impedes, although will not necessarily inhibit, fog formation.
A possible mechanism for fog formation within turbulent flows is identified, wherein regions of increased liquid water content form within the low-speed streaks of the near-wall cycle.
These streaks are energized in the moist cases due to reduced dissipation of turbulence kinetic energy compared to the dry case, although in both cases the streaks are less energetic and persistent than in neutrally stratified flow.

\vspace{0.5cm}

\textcopyright \hspace{0.02cm} 2019 California Institute of Technology. U.S.~Government sponsorship acknowledged.

\vspace{0.5cm}

\end{abstract}

%
\section{Introduction}
\label{sect:intro}

Fog is an important meteorological phenomenon that can have wide-reaching influences on human lives, transportation, and the economy \citep{Gultepe07}.
In the case of radiation fog, it forms within the stably stratified atmospheric boundary layer (SBL), wherein sufficient surface cooling leads to the air reaching its saturation point. Typically, this cooling occurs during nocturnal clear-sky conditions due to longwave radiation.
Fog is defined when the resulting suspension of water droplets reduces visibility to below 1 km, while visibilities above this threshold but below approximately 11 km are called mist \citep{NOAA17}.
Despite the importance and several decades of extensive research on fog,
accurate forecasts remain challenging \citep{Steeneveld15}. This is in part due to a lack of understanding
of some of the fundamental physical processes involved in fog formation and the early stages of its growth \citep{Gultepe07}.


\cite{Taylor17} provided a simple thermodynamic analysis of fog formation using a Clausius--Clapeyron diagram,
 in which cooling, mixing, and moistening are the three principle processes involved.
 All three processes may occur simultaneously during fog formation, although mixing alone is often not significant enough to cause fog \citep{Teixeira99}.
 Moreover,
 the role of turbulence and mixing on fog formation
 has often received  a range of interpretations \citep{Gultepe07}.
One hypothesis \cite[e.g.][]{Brown76,Roach76} is that very low, or absent, winds are necessary for
radiation fog to form, as otherwise the vertical mixing produced by turbulence draws drier, warmer air from aloft and prevents the air from reaching its saturation point.
An alternative view \cite[e.g.][]{Rodhe62,Welch86,Duynkerke99} is that turbulent mixing is essential, as it combines airmasses of different humidity and temperature 
such that saturation is achieved. 
Furthermore, both mechanisms might be responsible for fog formation depending on the specific conditions.
Here, we use direct numerical simulations (DNS) of a moist, stably stratified surface layer undergoing cooling as an idealized system  to study the fundamental mechanisms and relationship between 
turbulent mixing and fog formation.

Recently, large-eddy simulations (LES) of fog have had some success in simulating the main characteristics and qualitative behaviour 
of the fog life cycle in stably stratified environments \citep{Nakanishi00,Porson11,Bergot13}. Further LES studies \citep{Bergot16, Maronga17,Mazoyer17} have attempted to quantify
the effect of turbulence on fog formation and its life cycle. However, a core uncertainty here is the ability of LES to simulate stably stratified flows,
especially when the ground cooling is sufficient to lead to partial or complete turbulence collapse.
Ideally, LES resolves the largest, energy-containing eddies and uses a subgrid-scale (SGS) model to account for the dissipative actions of the smallest eddies.
In stably stratified flows, however, the largest eddies become suppressed due to the effort of drawing up heavier (cooler) fluid from below and pulling down lighter (warmer) fluid from above.
With increasing stability the largest energy-containing scales approach the grid size, potentially leading the SGS model to predict near-zero  turbulent fluxes \citep{Chung14les}.
Alternatively, some models artificially predict excessive mixing \citep{deRoode17}.
While there are SGS models that can account for the effects of buoyancy in the weakly stable regime \citep[e.g.][]{Lilly62,Deardorff80,Moeng84,BouZeid10,Chung14les}, 
there are still challenges when turbulence collapse is considered.

Further complications are introduced to considering fog when the non-linear mixing processes during fog formation occur at small length scales.
For example, the buoyancy length scale, $L_b=w_{rms}/N$, gives the level of suppression of vertical motions due to the stable stratification \citep{Stull88}, where $w_{rms}$ is the root-mean-square (r.m.s.) vertical velocity fluctuations and $N$ is the buoyancy frequency.
From the field observations of \cite{Price11}, fog occurred with $N\approx0.08$ s$^{-1}$ and the vertical velocity fluctuations were near zero with $w_{rms}\lesssim0.1$ m s$^{-1}$. This therefore leads to $L_b\sim 1.2$ m. 
 Similarly, the Ozmidov length, $L_{Oz}=\sqrt{\epsilon/N^3}$, gives the smallest scale influenced by buoyancy, where $\epsilon$ is the turbulence dissipation. For typical values of $\epsilon\sim 10^{-3}$ m$^2$ s$^{-3}$ then $L_{Oz}$ is of the same order as $L_b$.
 Recent high-resolution field observations have also noted substantial temperature gradients close to the ground during fog formation with $N\approx0.4$ s$^{-1}$ \citep{Izett19}, suggesting $L_B$ and $L_{Oz}$ may be even smaller.
Ultimately, this implies that LES, with horizontal grid spacings of several metres and at best vertical resolutions of 1 m \cite[e.g.][]{Bergot13,Maronga17}, may not properly resolve the small-scale mixing in fog formation.
 For this reason, we will use DNS of the moist SBL in the present study, which to our knowledge is the first time such  a technique has been used to study an idealized representation of fog formation.
DNS solves the Navier--Stokes equations, directly resolving the dissipative Kolmogorov length scale with no turbulence parametrization.

%
%
%


The structure of the dry SBL was described by \cite{Monin70} as follows.
The overall boundary layer, of height $\delta$, is split into two regions, with the lowermost region termed the surface layer. Here, Coriolis effects can be neglected and the thickness of the surface layer, $h$, is on the order of tens of metres.
The surface layer is further divided, wherein buoyancy forces can be neglected in the so-called dynamic sublayer, with thickness on the order of the Obukhov length, $L$.
The buffer layer ($z\ll L$) exists close to the ground and accounts for viscous effects in the case of a smooth surface, or roughness effects otherwise. 
Above the buffer layer, viscosity or roughness becomes irrelevant  and the only remaining length scale is the distance to the ground, $z$. 
This yields the familiar logarithmic mean velocity profile, as in neutrally stratified wall-bounded turbulence.
The flow remains turbulent so long as the ground cooling is sufficiently weak, such that $L$ is large, in the so-called weakly stable regime \citep{Mahrt99}.
However, under sufficiently strong ground cooling, $L$ becomes small and turbulence can collapse completely.
This results in a laminar flow with so-called runaway cooling \citep{vandeWiel07}, and is termed the very stable regime \citep{Mahrt99}.
%

The dry SBL has been simulated in a variety of configurations using DNS.
The surface layer can be simulated using an open channel flow of height $h$, driven by a constant pressure gradient.
Two alternative approaches are typically used, in which either a constant cooling flux is applied to the ground and the transient response is studied \citep[e.g.][]{Nieuwstadt05,Flores11},
or  a constant temperature difference is applied between the bottom and top boundaries yielding a statistically steady flow \citep[e.g.][]{GarciaVillalba11}. 
The Coriolis force is neglected and  low-level jets and other large-scale features of the SBL are not observed.
Simulations of the Ekman layer under stable stratification, meanwhile,  attempt to  represent the full SBL and include the Coriolis force \citep{Shah14,Ansorge14,Gohari17}.
While some differences between Ekman layers and open channel flows were reported in the outer-layer of the flow (where $z\sim h$),
\cite{Ansorge14} and \cite{Flores18} explicitly compared the near-wall region (buffer and logarithmic layers) of the two flows.
They showed that the logarithmic velocity profile and the turbulence kinetic energy (TKE) budget were comparable in this region, suggesting that the
outer-layer does not have a significant influence on the near-wall flow.

While the Reynolds numbers of DNS are relatively low, 
appropriate non-dimensionalization of turbulent flows often exhibits Reynolds number similarity scaling. 
In particular, \cite{Ansorge14} demonstrated that the velocity profiles, TKE budget, and intermittency factor do not vary significantly with Reynolds number in the neutrally stratified Ekman layer. Furthermore, the stably stratified surface-layer simulations of \cite{Flores11}  demonstrated that the time evolution of the total mass flux and density gradient at the ground were similar for different Reynolds numbers with matched cooling rates.
Similar validation will be performed here, which will enable extrapolation and comparison of the results of the present study at relatively low Reynolds numbers to those found in the atmosphere.

In this paper we perform DNS of both dry and moist
horizontally homogeneous open-channel flows,
in which
the flow is initialized from a dry turbulent neutrally stratified flow and
a constant cooling flux is then applied to the ground.
This represents the surface layer just
after sunset undergoing radiative cooling \citep[as in][]{Nieuwstadt05}, 
wherein the resulting condensation and small-scale turbulent mixing are studied in the context of fog formation.
The numerical procedure and simulation set-up are described and validated in Sect.~\ref{sect:code}.
 Particular attention is given in Sect.~\ref{ssect:fogsuitability} to how this system can be treated as an idealized representation of the early stages of fog formation.
Results are presented in Sect.~\ref{sect:results}, including an analysis of the fog development (Sect.~\ref{ssect:fogdev}) and of the turbulence collapse in dry and moist flows (Sect.~\ref{ssect:turbcollapse}). Finally, conclusions are offered in Sect.~\ref{sect:conc}.

 %
%




%
%
%
%

\section{Methodology}
\label{sect:code}
\subsection{Governing Equations and Simulation Set-up}
\label{ssect:goveqns}

We simulate a  horizontally homogeneous open-channel flow driven by a constant pressure gradient and an imposed ground-cooling flux.
This represents the same set-up as \cite{Nieuwstadt05} and \cite{Flores11}, although in addition we include moisture effects to study the influence of condensation of liquid water.
The incompressible Navier--Stokes equations with Boussinesq approximation are solved in Cartesian coordinates, along with transport equations for temperature, $T$, and water vapour mixing ratio, $q_v$. A bulk approximation for the liquid water mixing ratio, $q_l$ is used, yielding
\begin{align}
\mb{\nabla}\cdot \mb{u} &= 0, 	\label{eqn:mass}\\
\pf{\mb{u}}{t}+\mb{u}\cdot\mb{\nabla u} &= -\f{1}{\rho}\nabla \pi +\nu \nabla^2\mb{u}-\mb{i}\f{1}{\rho}\df{P}{x} + \mb{k}\mathcal{B}, 	\label{eqn:mmtm}\\
\pf{T}{t}+\mb{u}\cdot\mb{\nabla} T &= \f{L_v}{c_p} C_d + \nu_T \nabla^2 T,	\label{eqn:theta}\\
\pf{q_v}{t}+\mb{u}\cdot\mb{\nabla} q_v &= - C_d + \nu_v\nabla^2 q_v,	 \label{eqn:qv}\\
\pf{q_l}{t}+\mb{u}\cdot\mb{\nabla} q_l &= C_d + \nu_v\nabla^2 q_l, 	\label{eqn:ql}
\end{align}
where 
$\mb{u}=(u,v,w)$ is velocity in the streamwise (or mean flow, $x$), spanwise ($y$) and vertical ($z$) directions; 
$\pi$ is the pressure perturbation; 
$\rho$ is the constant air density;
$t$ is time; 
$\mb{i}$ and $\mb{k}$ are the unit vectors in the streamwise and vertical directions, respectively;
$\dd P/\dd x$ is the imposed constant streamwise pressure gradient that drives the flow;
$\nu$ is the constant molecular kinematic viscosity;
$\nu_T$ and $\nu_v$ are the thermal and water vapour molecular diffusivities, which are constant and related to viscosity through the molecular Prandtl number, $Pr_m=\nu/\nu_T=\nu/\nu_v=0.71$;
and
$C_d$ is the condensation rate,
which maintains thermodynamic equilibrium with zero supersaturation \citep{Grabowski90}.
$\mathcal{B}$ is the buoyancy, defined as
\begin{equation}
\mathcal{B} \equiv g\left[\f{T-T_0}{T_0}+\epsilon(q_v-q_{v0})-q_l\right],
\end{equation}
where $g=9.81$ m s$^{-2}$, $T_0$ and $q_{v0}$ are the reference temperature and water vapour mixing ratio, respectively, and $\epsilon+1=R_v/R_d\approx1.61$ is ratio of the gas constants for water vapour and dry air.
The latent heat of condensation is $L_v=2.5\times 10^6$ J kg$^{-1}$ and specific heat at constant pressure is $c_p=1005$ J kg$^{-1}$ K$^{-1}$.

The governing equations above are solved using the finite volume, non-hydrostatic anelastic model EULAG, broadly documented in the literature \citep{Grabowski90,Smolarkiewicz97,Smolarkiewicz98,Grabowski02,Andrejczuk04,Kurowski14}, with a review in \cite{Prusa08}.
 The second-order accurate Eulerian (flux form) mode of EULAG is used in the present work.
The present bulk approximation for $q_l$, along with a more detailed microphysical scheme that models supersaturation and the size dependence of multiple cloud droplets on sedimentation and evaporation, have previously been tested with EULAG in \cite{Andrejczuk04,Andrejczuk06,Andrejczuk09}.
There, the authors performed DNS of cloud filaments embedded within decaying turbulence in a triply periodic domain. 
They only observed moderate differences between the bulk and detailed microphysics schemes,
indicating that the bulk liquid water scheme used for the present DNS of the moist SBL is likely adequate.
A similar bulk scheme has also been used in DNS studies of stratocumulus cloud tops \citep{Mellado10evap,Mellado10,Mellado14}
as well as in the LES study of radiation fog in  \cite{Nakanishi00}.


Simulations are conducted in an open channel, consisting of periodic boundary conditions in the lateral directions.
The lower wall (or ground, subscript $g$) at $z=0$ consists of no-slip ($u=v=0$) and impermeability ($w=0$) conditions
 with an imposed  constant (cooling) heat flux ($H_g<0$, described in Sect.~\ref{ssect:nondim}) and zero total water flux ($\dd (q_v+q_l)/\dd z=0$).
The upper boundary at $z=h$ is a free-slip ($\dd u/\dd z=\dd v/\dd z=0$) impermeable  wall with zero total water flux and constant temperature $(T-T_0=0)$.
 Both dry (only Eqs.~\ref{eqn:mass}--\ref{eqn:theta}, with no water vapour and liquid water) and moist (Eqs.~\ref{eqn:mass}--\ref{eqn:ql}) flows are simulated
 with the same imposed ground heat flux, so that the effects of liquid water can be analyzed independently of the ground cooling.

A snapshot from the turbulent quasi-stationary neutrally stratified dry case is used to initialize the flow, after which the ground cooling flux is applied.
 The initial thermodynamic state is set to air at standard atmospheric pressure (1013 hPa) with density $\rho=1.265$ kg m$^{-3}$ and viscosity $\nu=1.38\times 10^{-5}$ m$^2$ s$^{-1}$.
The temperature is initialized as $T_0=279.15$ K everywhere, although a case with higher temperature of $T_0=285.15$ K is also simulated and shows little difference (see Appendix 2).
The initial relative humidity is set to  $RH_0=99.9$\% throughout the domain, corresponding to $q_{v0}\approx 6.24$ g kg$^{-1}$. 
This large value of $RH_0$ is used as lower values are computationally inefficient at reaching saturation,
and is discussed further in Sect.~\ref{ssect:fogsuitability}.
The air is initialized with no liquid water, $q_{l0}=0$.

In subsequent sections,
 temporal averaging is  denoted with an over bar, $\overline{\cdot}$, while spatial averaging in the horizontal plane is denoted by angled brackets, $\langle \cdot\rangle$.
Velocity fluctuations are defined based on the difference between the instantaneous, spatially dependent velocity  and its spatially averaged velocity at a given vertical location and time,
$u'(x,y,z,t)=\langle{u}\rangle(z,t) - u(x,y,z,t)$. Root-mean-square (r.m.s.) velocity fluctuations are then defined as $u'_{rms}(z,t) = \langle{u'^2}\rangle^{1/2}$.

\subsection{Applicability to Fog Formation}
\label{ssect:fogsuitability}

The simulation set-up detailed above enables us to study the influence of moisture on the stable surface layer.
More generally however, this system can also be treated as an idealized representation of fog formation within the atmosphere.
Here, we are isolating just the influence of the small-scale turbulent mixing on fog formation over the relatively short turbulence time scales.
As such we deliberately neglect other competing processes;
some of the key assumptions and limitations of this idealized system in the context of fog formation are discussed below.

The present bulk liquid water approximation assumes that the disperse liquid phase, $q_l$, can be modelled as a continuum. As discussed in \cite{Mellado10}, this condition is generally not met at the tops of stratocumulus clouds, as the cloud droplets of diameter $d\approx 10$ $\mu$m and number density $N_d=1000$ cm$^{-3}$ are too sparse within a volume of the order of the Kolmogorov length scale, $l_\eta=1$ mm (i.e., that used in DNS). A similar problem exists for fog, as the Kolmogorov length scale is of the same order, however the droplets are even sparser with $N_d\approx10$--100 cm$^{-3}$ and diameter 1--10 $\mu$m \citep{Roach76,Price11}. 
In addition, we  assume the liquid water diffusivity is equal to that of water vapour.  We also assume 
 thermodynamic equilibrium, wherein phase changes occur instantaneously to maintain saturation through the condensation rate $C_d$. 
These assumptions may not hold for fog formation and, furthermore, predicting their impact on the present results is not trivial.
However, they form useful approximations which are commonly employed in  DNS studies of cloudy boundaries, notably 
stratocumulus cloud tops \citep[e.g.,][]{Mellado10evap,Mellado14,deLozar15,deLozar17}.
As this is a dynamically similar problem to fog, which involves the mixing of saturated and unsaturated air within a stably stratified environment,
we employ the same assumptions in the present study.

Direct numerical simulation of  the top of stratocumulus clouds have also shown radiation and droplet sedimentation to be an important process governing the mixing of clear and cloudy air \citep{deLozar15,deLozar17}.
While stratocumulus cloud tops are a dynamically similar problem, fog in its early formation stages is unlikely to be optically thick enough for radiation effects to be significant. It is only once the fog is a few metres thick (i.e. on the scale of the present domain) that radiation is thought to become important \citep{Oliver78}, wherein the radiative cooling enhances fog growth. In the present simulations the liquid water mixing ratio is typically less than 0.01 g kg$^{-1}$ and so we neglect radiation effects. 
For simplicity, we also neglect droplet sedimentation effects, similar to the DNS study of stratocumulus tops by \cite{Mellado14}, which only considered shear and evaporative cooling in isolation. 
Moreover, it has been suggested that droplet sedimentation and radiative cooling are strongly dependent on one another, based on simple one-dimensional fog models \citep{Brown76,Bott90} and studies of individual cloud or fog droplets \citep{Roach76droplet,Barkstrom78}.
This suggests that both processes would need to be modelled together.

The present simulations  neglect any moisture flux through the ground and more broadly there are no soil-vegetation effects, which are important in fog formation \citep{Gultepe07}.
Additional simulations were run with an imposed downwards moisture flux, representing  dew deposition and hygroscopic absorption.
The magnitude of this moisture flux was approximately equivalent to 20 g m$^{-2}$ hr$^{-1}$, similar to that observed during one of the fog formation cases in the field study of \cite{Price14}.
However, in the present system this caused a significant drying influence and completely inhibited all condensation.
Similarly, reducing the initial relative humidity to a uniform value of 99.5\% (with zero moisture flux boundaries) to enable more realistic water vapour gradients to develop also inhibited condensation. 
These two results are  presumably due to a lack of moisture entering the system, so that the ground cooling alone is then insufficient to lead to condensation within reasonable simulation times.
The present simulations therefore use the simpler configuration of zero moisture flux and a high initial relative humidity,
rather than attempting to specify the complex  moisture fluxes that occur with the ground or the atmosphere  (either due to horizontal heterogeneity or from higher up in the boundary layer) at such small scales necessary for DNS.


\subsection{Scaling Variables}
\label{ssect:nondim}
The characteristic velocity scale within the surface layer is the friction velocity \citep{Monin70}.
Following \cite{Flores11}, we distinguish between the (constant) friction velocity obtained from the imposed driving pressure gradient,  $U_\star^2 = -(h/\rho)\dd P/\dd x$,
and from the spatially averaged wall-shear stress, $u_\tau^2(t)=\tau_w/\rho=\nu \dd  \langle u\rangle/\dd z|_g$.
 The   latter can vary with time
  due to accelerations in the bulk velocity, $u_b(t) = (1/h)\int_0^h \langle u\rangle \id z$, caused by the ground cooling.
For the neutrally stratified dry cases, a statistically steady state is obtained such that the bulk velocity 
 is approximately constant over time and $\overline{u_\tau}=U_\star$. 
 As such, $U_\star$ can be regarded
 as the reference friction velocity of the neutrally stratified case, which is matched for all simulations with the same $Re_\star$.

The characteristic temperature, water vapour, and length scales will be related to $H_g<0$,  the imposed (cooling) heat flux that is applied to the ground at $t=0$.
This heat flux is matched between respective dry and moist cases.
 In response, sensible and, when saturation occurs, latent heat fluxes  will be directed toward the surface.
These fluxes, when horizontally averaged, are given as \citep{Monin70}
\begin{eqnarray}
H_s(z,t) = \rho c_p \left(\nu_T \df{\langle T\rangle}{z}- \langle w' T'\rangle\right), \\
H_l(z,t) = \rho L_v \left( \nu_v \df{\langle q_v\rangle}{z} - \langle w' q_v'\rangle\right).
\end{eqnarray}
In meteorology, the conduction (gradient) terms of temperature and water vapour are often ignored as they are considered negligible relative to the turbulent fluxes.
However, as we are resolving the viscous sublayer with impermeability constraint in the present DNS, $w(z=0)=0$, the gradient terms become significant at the ground \citep{Businger82} whereas the turbulent fluxes vanish.
As such, the imposed ground heat flux can be seen to prescribe the temperature and water vapour gradients, 
\begin{eqnarray}
H_g &=& H_s(0,t)+H_l(0,t) \\
       &=&\rho c_p \nu_T \left.\df{\langle T\rangle}{z}\right|_g + \rho L_v \nu_v \left.\df{\langle q_v\rangle}{z}\right|_g. \label{eqn:Hg}
\end{eqnarray}

In the SBL, the characteristic length scale is taken to be the Obukhov length, $L$, as discussed in Sect.~\ref{sect:intro}.
This can be interpreted as the height at which buoyancy effects dominate over mechanical (shear) production of turbulence kinetic energy \citep{Stull88}.
The Obukhov length is a function of the ground cooling and is defined as
\begin{equation}
\label{eqn:obukhov}
L = -\f{U_\star^3/\kappa}{(g/T_{0}) H_g/(\rho c_p)},
\end{equation}
where $\kappa=0.41$ is the von K{\'a}rm{\'a}n constant, included for historical reasons.
In LES models and field studies the heat flux at the ground is often taken to be the vertical turbulent heat flux, 
$\langle w'T'\rangle$, at the lowest level.
However, as mentioned above, these fluxes vanish at the ground
so that Eq.~\ref{eqn:Hg} is used to define $H_g$.
The ratio between the surface layer height (or channel half height) and the (imposed) Obukhov length, $h/L$, is often reported in stably stratified DNS studies; 
when $h/L\lesssim1$ the cooling is relatively weak while $h/L\gtrsim1$ corresponds to stronger cooling.
Note, however, that the ratio $h/L$ is not a suitable measure for turbulence collapse as it is Reynolds number dependent \citep{Nieuwstadt05,Flores11}.
In addition, there are two characteristic length scales  in neutrally stratified wall turbulence, namely $h$, the outer-layer or large-scale length scale and $\nu/U_\star$, the inner-layer, near-ground or viscous length scale.
Both of these will be used here; superscript $+$ indicates non-dimensionalization on $\nu$ and $U_\star$. For example, $z^+= zU_\star/\nu$ is the vertical position non-dimensionalized on the viscous length scale.

\setlength{\unitlength}{1cm}
\begin{figure}
\centering
  \includegraphics[trim=0 0 20 0,clip=true]{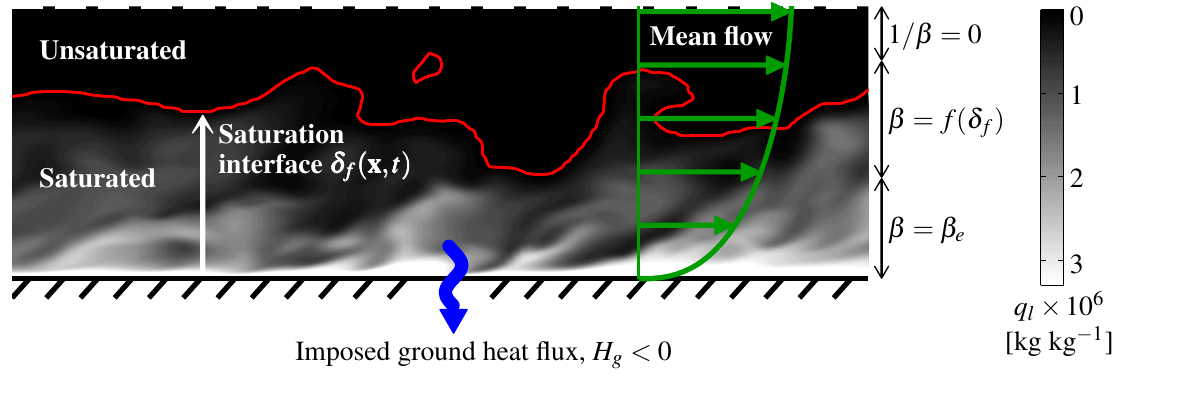}
  \vspace{-0.3cm}
\caption{
Cross-section of the streamwise--vertical plane showing an instantaneous snapshot of the liquid water mixing ratio for case 395M06.
The saturation interface, $\delta_f$, where $q_l>0$, varies in space and grows in time
as the ground continues to be cooled through the imposed heat flux, $H_g$.
Variations in the Bowen ratio, $\beta$, are shown on the right
}
\label{fig:fogSketch}
\end{figure}

The horizontally averaged Bowen ratio, $\beta(z,t)=H_s/H_l$, gives the ratio between the sensible and latent heat fluxes.
At a given vertical height, $z$, there could exist  regions of unsaturated air ($H_l=0$) alongside  regions of non-zero liquid water mixing ratios ($H_l >0$).
Hence, as with the heat fluxes, the Bowen ratio is a function of both vertical position and time. 
When a vertical level is completely saturated, $\beta$ will tend towards the equilibrium Bowen ratio $\beta_e$ due to the use of the bulk condensation model that maintains thermodynamic equilibrium (Sect.~\ref{ssect:goveqns}). Based on the present thermodynamic state, this can be estimated as $\beta_e=(c_p/L_v)/(\partial q_{sat}/\partial T)\approx0.92$ \citep{Stull88}.
Meanwhile, in the dry cases and moist cases where the air is completely unsaturated at a given $z$, there is no latent heat flux and $1/\beta=0$ (cf.~Fig.~\ref{fig:fogSketch})

The instantaneous friction temperature and water vapour can now be defined as
 \begin{eqnarray}
 \label{eqn:thtau}
 T_\tau(z,t) &=& -\f{H_s}{\rho c_p u_\tau} = -\f{H_g/(1+1/\beta)}{\rho c_p u_\tau},\\
  \label{eqn:qvtau}
 q_\tau(z,t) &=& -\f{H_l}{\rho L_v u_\tau} =  -\f{H_g/(\beta+1)}{\rho L_v u_\tau}.
  \end{eqnarray}
Crucially, these can not be known a priori as they depend on the evolution of $\beta(z,t)$ (or the saturation interface $\delta_f$) and instantaneous wall-shear stress through $u_\tau(t)$.
These quantities vary with height and time, which can be interpreted using the internal boundary layer framework \citep{Elliott58,Panofsky64}.
For example, after a step change in roughness, the flow above the newly formed internal boundary layer (IBL) still depends on the  friction velocity of the original surface upstream,
while that within the IBL scales with the friction velocity of the new surface.
In the present moist simulations, the flow at the top of the domain when unsaturated would depend on $T_\tau$ defined with $1/\beta=0$, as the information regarding the saturated condition has not yet reached this height.
The flow close to the ground, where liquid water has condensed, would meanwhile depend on $T_\tau$ defined with $\beta\rightarrow\beta_e\approx0.92$.
Note that rather than growing spatially, as in conventional IBL depictions, the saturation interface grows temporally due to the periodic boundary conditions in the streamwise direction, which imposes horizontal homogeneity \citep{Mellado12,Kozul16}.
We can define a constant friction temperature and water vapour, analogous to the $U_\star$ friction velocity, using the  equilibrium Bowen ratio,
 \begin{eqnarray}
 \label{eqn:thstar}
 T_\star &=& -\f{H_{s,g}}{\rho c_p U_\star} = -\f{H_g/(1+1/\beta_e)}{\rho c_p U_\star},\\
  \label{eqn:qvstar}
 q_\star &=& -\f{H_{l,g}}{\rho L_v U_\star} =  -\f{H_g/(\beta_e+1)}{\rho L_v U_\star};
  \end{eqnarray}
  however this assumes fully saturated conditions across the entire domain. The value of $T_\tau$ and $q_\tau$ would therefore converge to $T_\star$ and $q_\star$ (for all $z$), although only when $t$ becomes sufficiently large.

\subsection{Description of Cases}
\label{ssect:sims}

\begin{table}
\caption{Details of the simulations performed.
$N_{runs}$ is the number of runs using unique initialization snapshots from the dry, neutrally stratified case
}
\label{tab:sims}   
\centering
\begin{tabular}{ l l l l l l l}
\hline\noalign{\smallskip}
 ID			& $Re_{\star}$		& $N_{runs}$		& $h/L$	& Final state\\
\noalign{\smallskip}\hline\noalign{\smallskip}
395D00			& 395			& 1		& 0		& Turbulent	\\
395D041		& 395			& 1		& 0.41	& Turbulent	\\
395M041		& 395			& 1		& 0.41	& Turbulent	\\
395D06		& 395			& 3		& 0.6		& Turb., Lam.	\\
395M06		& 395			& 3		& 0.6		& Turbulent	\\
395D07		& 395			& 1		& 0.7		& Laminar	\\
395M07		& 395			& 1		& 0.7		& Laminar	\\
395D205		& 395			& 1		& 2.05	& Laminar	\\
395M205		& 395			& 1		& 2.05	& Laminar	\\
590D06		& 590			& 1		& 0.6		& Turbulent	\\
590M06		& 590			& 1		& 0.6		& Turbulent	\\
590D085		& 590			& 1		& 0.85	& Turbulent	\\
590M085		& 590			& 1		& 0.85	& Turbulent	\\
\hline
\end{tabular}
\end{table}

Table \ref{tab:sims} details the different simulations conducted.
We independently vary the friction Reynolds number, $Re_\star=U_\star h/\nu$ and Obukhov length, $L$.
Cases are referred to with an ID of their friction Reynolds number, whether the case is dry or moist, and the ratio between channel half height and Obukhov length. For example,
case 395M06 is a moist case performed at $Re_\star=395$ with $h/L=0.6$.
A constant grid-spacing is used in the horizontal directions, with $\Updelta x^+=\Updelta x U_\star/\nu\approx9.7$ and $\Updelta y^+\approx4.8$.
A hyperbolic tangent grid stretching is used in the vertical direction \citep{Moin82}, resulting in grid spacings at the ground
and channel centre of $\Updelta z|_{z=0}^+\approx 0.35$ and $\Updelta z|_{z=h}^+\approx 7.0$, respectively.
These grid spacings are in good agreement with those used in previous stably stratified DNS studies  \cite[e.g.][]{Flores11,GarciaVillalba11,Gohari17}.
The DNS studies of \cite{Ansorge14} and \cite{Shah14}, meanwhile, had matched streamwise and spanwise grid spacings with $\Updelta x^+=\Updelta y^+\gtrsim4$ and thus had a finer streamwise grid spacing than the present case. However,
if we assume that viscous dissipation equals the production of total turbulence kinetic energy, $\epsilon = -u_b \dd P/\dd x$ \citep[e.g.][]{Nieuwstadt05},
then the present grid spacings can be related to the Kolmogorov length, $\eta=(\nu^3/\epsilon)^{1/4}$, as $\Updelta x = 4.0 \eta$, $\Updelta y = 2.0\eta$, $\Updelta z|_{z=0}=0.15\eta$ and $\Updelta z|_{z=h}=2.9\eta$.
They are therefore all $O(\eta)$, in agreement with the grid-spacing recommendations for conventional DNS \citep{Moin98}.
For the cases with $Re_\star=395$ and $h/L=0.6$, multiple runs are performed in which unique initialization snapshots from the dry neutrally stratified case are used for both dry and moist cases (395D06 and 395M06).
The computational domain size is set to $L_x\times L_y=2\pi h\times \pi h$, which is commonly employed for neutrally stratified dry simulations \citep{Lozano14,Munters16}.
The effect of the domain size is investigated in Appendix 1, showing that while the domain is relatively small for stably stratified flows, it should not alter the conclusions of this paper.
The present simulations use $256\times256\times128$ grid points for the $Re_\star=395$ cases and $384\times384\times192$ grid points for the $Re_\star=590$ cases to obtain the grid spacings mentioned above.




\subsection{Validation of EULAG}
\label{ssect:valid}


\setlength{\unitlength}{1cm}
\begin{figure}
\centering
  \includegraphics{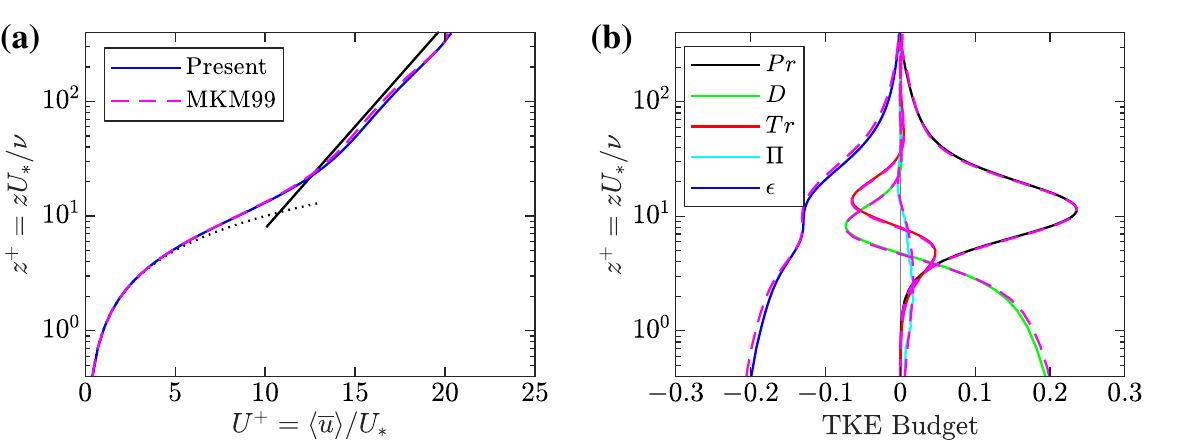}\\
  \vspace{-0.05cm}
\caption{Profiles of (a) velocity and (b) TKE budget for 
the present dry neutrally stratified case (solid lines)
and the DNS data of \cite{Moser99} (magenta dashed lines) at $Re_\star=h U_\star/\nu= 395$.
In (a), 
the dotted black line shows the viscous sublayer velocity, $U^+=z^+$,
while the solid black line shows the logarithmic velocity profile, $U^+=(1/0.41)\log(z^+)+5.2$.
The TKE budget terms in (b) are given in Eq.~\ref{eqn:tkebudget} and normalized on $U_\star^4/\nu$;
note that the buoyancy production, $B$, is zero
}
\label{fig:dryvalid}
\end{figure}

The neutrally stratified dry case, 395D00, is first validated with the DNS data of \citet[][herein MKM99]{Moser99}
at a friction Reynolds number $Re_\star=hU_\star/\nu=395$.
This case obtains a statistically steady state independent of the initial conditions, identified by a linear profile of the total stress profile, $-\langle\overline{u'w'}\rangle+\nu\dd\langle\overline{u}\rangle/\dd z$ \citep{Kim87}.
The flow is then temporally averaged over $tU_\star/h\approx20$ large-eddy turnover times.
Figure \ref{fig:dryvalid} shows the mean velocity profile
and 
(resolved) TKE budget for the present data from EULAG, along with the DNS data of MKM99.
Good agreement is observed between the two datasets, indicating that EULAG can be readily used for surface-layer simulations.
The budget for the horizontally averaged TKE, $e=\langle u_i' u_i'\rangle/2$, is shown in Fig.~\ref{fig:dryvalid}b, where the terms correspond to
\begin{eqnarray}
\hspace{-0.25cm}\pf{ e}{t} \hspace{0.1cm}= &\hspace{-0.75cm}
\overbrace{-\langle u' w' \rangle \pf{\langle u\rangle}{z}}^{Pr} \hspace{0.15cm}+\hspace{0.15cm} 
\overbrace{\nu \pfsq{e}{z}}^{D}  \hspace{0.15cm}+\hspace{0.15cm} 
\overbrace{-\pf{\langle e w'\rangle}{z}}^{Tr}\nonumber\\ 
&\hspace{-1.0cm}+\hspace{0.15cm} 
\underbrace{-\f{1}{\rho}\pf{\langle \pi'w'\rangle}{z}}_{\Pi} \hspace{0.15cm}+\hspace{0.15cm} 
\underbrace{\f{g}{T_0}\langle w'T'\rangle}_{B} \hspace{0.15cm}-\hspace{0.15cm} 
\underbrace{\nu\left\langle\pf{u_i'}{x_j}\pf{u_i'}{x_j}\right\rangle}_{\epsilon},
\label{eqn:tkebudget}
\end{eqnarray}
where $Pr$ is the mechanical (or shear) production, $D$ the viscous diffusion, $Tr$ the turbulent transport, $\Pi$ the pressure correlation, $B$ the buoyant production and $\epsilon$ the (pseudo-) dissipation. 
The left hand side, referred to as the tendency or residual, is zero for this neutrally stratified flow due to it achieving a statistically steady state.
Conventional summation notation is used for repeated subscripts of $i$ and $j$.
The Reynolds stress terms, $\langle\overline{u_i'u_j'}\rangle$ (not shown) were also in good agreement with MKM99, although the vertical Reynolds stress, $\langle\overline{w'w'}\rangle$ went to zero at $z=h$ due to the use of the slip, impermeable boundary condition. This reduction of $\langle\overline{w'w'}\rangle$ began at approximately $z/h\approx0.8$ in agreement with other open channel simulations \citep[e.g.][]{MacDonald17}.

\section{Results}
\label{sect:results}
\subsection{Temporal Evolution of Turbulent Fluctuations}
\label{ssect:timehist}

\setlength{\unitlength}{1cm}
\begin{figure}
\centering
  \includegraphics{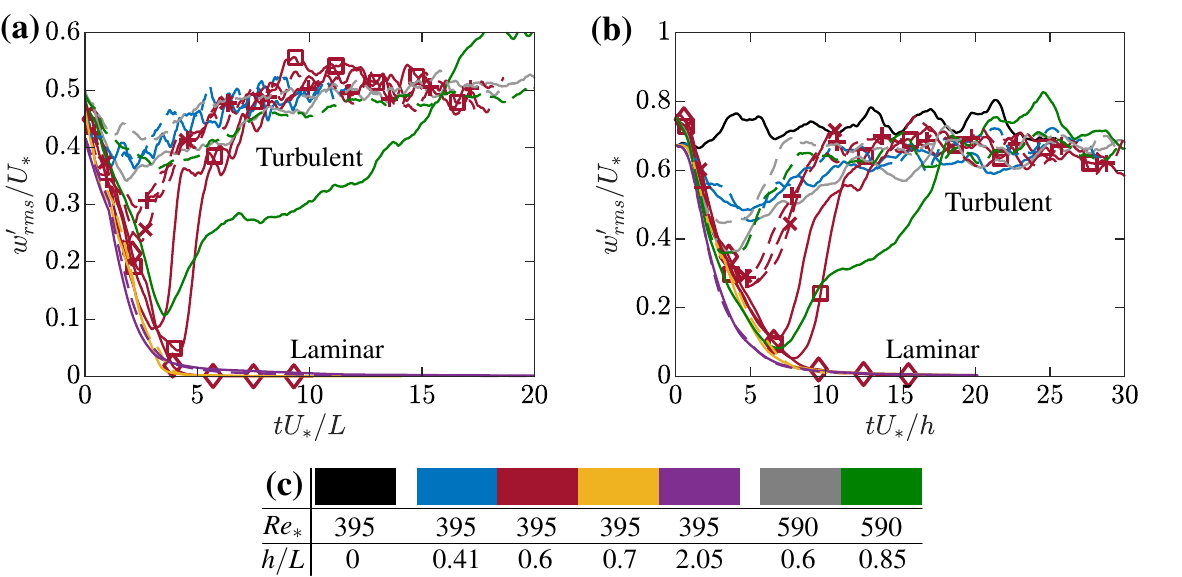}\\
  \vspace{-0.05cm}
\caption{Time series of 
 the r.m.s.~vertical velocity   fluctuations at 
 (a) $z^+ \approx 15$,
and at (b) $z/h\approx2/3$.
Line colours defined in (c), corresponding to cases listed in Table \ref{tab:sims}.
Solid lines represent dry cases, while dashed lines represent moist cases.
Additional  runs for $Re_\star=395$, $h/L=0.6$ with unique initialization snapshots are indicated with 
symbols;
\protect\raisebox{0.2ex}{\protect\scalebox{1.3}{$\color{myred}\boldsymbol{{\diamond}}$}}
and
\protect\raisebox{0.3ex}{\protect\scalebox{1.0}{$\color{myred}\boldsymbol{\pmb{\times}}$}}
are dry and moist cases, respectively, for the same snapshot; 
likewise
\protect\raisebox{0.3ex}{\protect\scalebox{0.8}{$\color{myred}\boldsymbol{\pmb{\pmb{\square}}}$}}
and
\protect\raisebox{0.4ex}{\protect\scalebox{0.9}{$\color{myred}\boldsymbol{\pmb{+}}$}}
for the other snapshot.
Note the time normalization is different in (a) and (b) following \cite{Flores11};
as such the neutrally stratified case ($L=\infty$) is not plotted in (a)
}
\label{fig:timehist_wd}
\end{figure}

The imposed ground cooling can either lead to complete turbulence collapse, resulting in laminar flow (very stable regime), or 
maintenance of the turbulent flow (weakly stable regime). To highlight these two regimes, Fig.~\ref{fig:timehist_wd} shows the time series of the r.m.s.~of the vertical velocity fluctuations
at two different heights, $z^+\approx 15\approx 0.04 h^+$ (i.e. close to the ground, within the buffer layer) and $z/h\approx2/3$.
A moving average filter of size 1$h/U_\star$ is applied to the present time series for clarity.
 Figure \ref{fig:timehist_wd} matches Fig.~3 of \cite{Flores11} and,
 as discussed in \cite{Flores11}, the temporal adjustment to the imposed cooling scales with 
$L/U_\star$ in the buffer layer (Fig.~\ref{fig:timehist_wd}a),
while in the outer layer  (Fig.~\ref{fig:timehist_wd}b) it scales with $h/U_\star$.
The neutrally stratified case is therefore not shown in Fig.~\ref{fig:timehist_wd}a as $L=\infty$.

In general, the cooling causes a reduction in the vertical velocity fluctuations within the first 5--10 large-eddy turnover times. 
In the weakly stable cases 
the turbulence then recovers to magnitudes close to the neutrally stratified case, 
while for very stable cases 
the turbulence collapses and the fluctuations tend to zero.
As will be seen later, in the moist cases (dashed lines) saturation begins at the ground almost immediately. The associated latent heat release and buoyancy effects due to condensation   appear to cause increased mixing in the moist case, leading to the turbulent fluctuations remaining larger compared to the dry cases (solid lines) at matched $h/L$. This is especially evident when the flow is close to laminarization.
In particular, for cases 395D06 and 395M06 three different initialization snapshots were run with cooling of $h/L=0.6$, wherein the cooling is sufficient to lead to complete turbulence collapse for one of the dry runs (red solid line with diamond symbols) but not the other two. Meanwhile, all three moist runs remained weakly stable and maintained turbulence.

The exact critical value of $h/L$ that leads to turbulence collapse has been recognized as being
Reynolds number dependent,
with \cite{Nieuwstadt05} observing a critical value of $h/L =0.51$ for $Re_\star\approx360$, while
\cite{Flores11} observed the critical value to be $h/L =0.82$ for $Re_\star\approx 560$.
The present values of $h/L\approx0.6$ for $Re_\star\approx395$ and $h/L\gtrsim0.85$ for $Re_\star\approx590$ for the dry cases therefore agree with this trend.
Due to the Reynolds number dependence in $h/L$, \cite{Flores11} suggested that
$L_\tau u_\tau/\nu$ is a better measure for the critical cooling level, where 
$L_\tau$ is the Obukhov length defined using the instantaneous wall-shear stress $u_\tau$ (as opposed to the constant driving $U_\star$, as in Eq.~\ref{eqn:obukhov}).
The two dry runs of 395D06 which sustained turbulent flow were observed to have a minimum $L_\tau u_\tau/\nu\approx95$, 
while the other initialization snapshot, which lead to turbulence collapse, had $L_\tau u_\tau/\nu\le75$.
This agrees with the observation in \cite{Flores11} that turbulence collapses when $L_\tau u_\tau/\nu\lesssim 100$,
due to insufficient scale separation between 
the length scale of turbulent production in the buffer region (approximately $100\nu/u_\tau$)
and
the buoyancy length scale ($L_\tau$).
The moist cases with $Re_\star = 395$ and $h/L=0.6$ had a minimum $L_\tau u_\tau/\nu\approx225$ for all three initialization snapshots, although
 for $h/L =0.7$ it was $L_\tau u_\tau/\nu\approx70$ and led to turbulence collapse.
 Compared to the dry case with critical $h/L =0.6$, the production of liquid water  enables a slightly larger
 cooling rate to be achieved before leading to turbulence collapse, with critical $h/L$ in the range 0.6--0.7.

 Note that, as discussed in \cite{Flores11} and \cite{GarciaVillalba11}, 
the asymptotic laminar friction Richardson number, $Ri_{\tau,l}=(2/\kappa)(h/L)Re_\star Pr_m$ for all the present cases is less than the  linear stability limit \citep{Gage68}. Therefore we would expect the cases with laminar flow to eventually recover a turbulent state given enough time.
Also, purely laminar flow across the entire surface layer is unlikely to occur in the real atmosphere due to 
turbulence generated by large-scale structures that are not captured in the present idealized system,
such as
 low-level jets and the breaking of gravity waves \citep{Flores11}.


 The effects of the relatively small computational domain size are evident in the dry higher Reynolds number cases (particularly 590D085, green solid line), where it takes longer to return to the statistically steady turbulent state. This is due to the existence of `locked' turbulent structures \citep{Flores11,GarciaVillalba11},
where adjacent turbulent and laminar patches become locked in place as a result of the periodic boundary conditions (see also Appendix 1).
The moist case, meanwhile, does not exhibit this behaviour for the same cooling rate. This is possibly due to the latent heat released during condensation, which would enhance mixing due to buoyancy production. 
Moreover, a positive feedback system exists wherein any tendency toward a laminar state with reduced mixing would result in enhanced, or runaway, cooling \citep{vandeWiel07}. This would therefore promote condensation-induced mixing and thus avoid the spatially locked laminar patches associated with turbulence collapse.
Note that this effect may be sensitive to the domain size \citep{GarciaVillalba11,Ansorge14} and would require further investigation.

\setlength{\unitlength}{1cm}
\begin{figure}
\centering
  \includegraphics{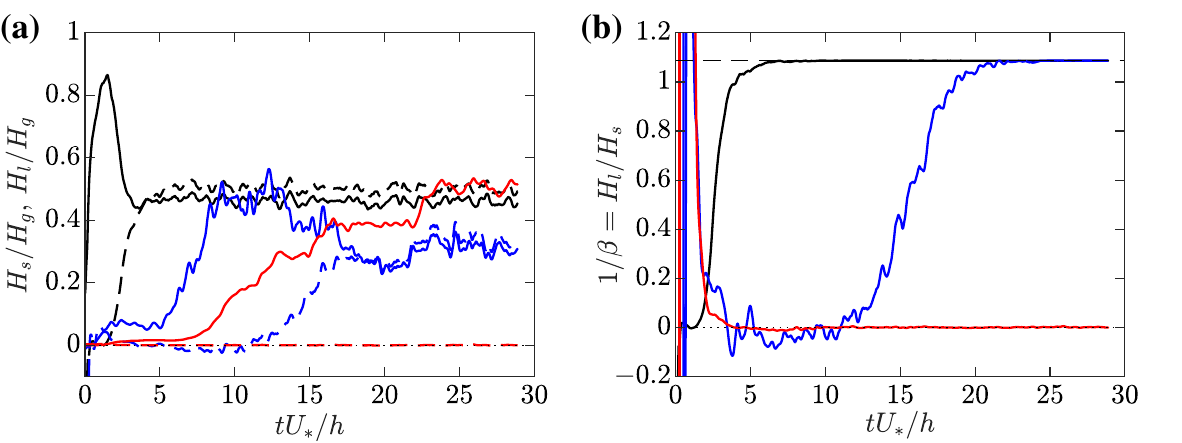}\\
    \vspace{-0.05cm}
\caption{
Time series of
(a) sensible (solid)
and
latent (dashed)
heat fluxes
normalized on the imposed ground heat flux, $H_g$,
and
(b) inverse of the Bowen ratio, $1/\beta=H_l/H_s$.
These are shown at
$z^+\approx 15$ (black),
$z/h=2/3$ (blue),
and $z=h$ (red)
for case 395M06.
Equilibrium Bowen ratio $1/\beta_{e}\approx1.09$ shown by horizontal dashed line in (b)
}
\label{fig:timehist_flux_beta}
\end{figure}

Before analyzing variations in the r.m.s.~of temperature fluctuations, Fig.~\ref{fig:timehist_flux_beta}a shows
how the sensible and latent heat fluxes transported by the fluid
change over time for case 395M06.
Close to the ground, at $z^+\approx 15$, the sensible heat flux  (black solid line) initially increases rapidly
due to the imposed ground heat flux. However, soon thereafter the temperature reaches saturation point
and liquid water condenses, resulting in an increase in the latent heat flux  (black dashed line) and decrease in the sensible heat flux.
Within five large-eddy turnover times, the latent and sensible heat fluxes reach
a statistically steady state
and the inverse of the Bowen ratio (Fig.~\ref{fig:timehist_flux_beta}b) is exactly equal to
its equilibrium value (horizontal dashed line).
This does not vary with time due to the use of the bulk condensation model,
which maintains equilibrium conditions.
A similar effect occurs for the heat fluxes higher up at $z/h=2/3$ (blue lines),
although this is delayed as the interface between saturated and unsaturated air requires time to grow upwards
(see sketch in Fig.~\ref{fig:fogSketch}).
Note that the sum of $H_s$ and $H_l$ at this height does not equal $H_g$ as the flow has not achieved a true
statistically steady state; this would only occur when the heat flux at the top of the domain
balances the imposed ground heat flux.
No significant amount of liquid water reaches the top of the domain ($z=h$, red line) so the latent heat flux  and Bowen ratio remains essentially zero.

\setlength{\unitlength}{1cm}
\begin{figure}
\centering
  \includegraphics{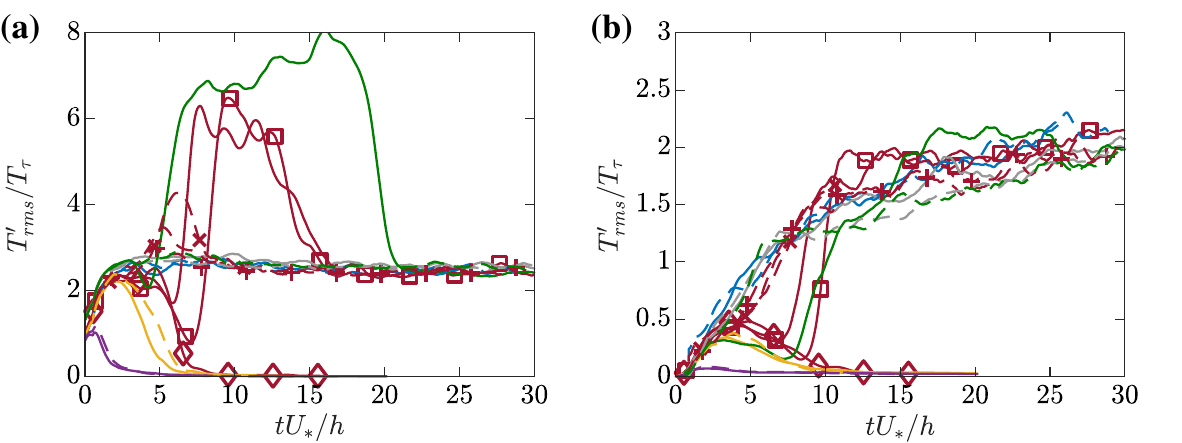}\\
  \vspace{-0.05cm}
\caption{
Time series of r.m.s.~temperature fluctuations at
(a) $z^+ \approx 15$,
and at (b) $z/h=2/3$.
Line styles same as Fig.~\ref{fig:timehist_wd}.
Temperature normalized on $T_\tau(z,t)$ (Eq.~\ref{eqn:thtau}), defined on instantaneous $\beta$ and $u_\tau$)
}
\label{fig:timehist_th}
\end{figure}

The r.m.s.~temperature fluctuations (Fig.~\ref{fig:timehist_th})
are initially zero before increasing with time, as $T$ is uniformly set to $T_0$ at $t=0$.
The temperature in Fig.~\ref{fig:timehist_th} is normalized on the
instantaneous, vertically dependent friction temperature $T_\tau(z,t)$ from Eq.~\ref{eqn:thtau}.
This choice of normalization results in good agreement between dry and moist turbulent cases with different cooling rates,
as well as for the different Reynolds number cases.
Figure \ref{fig:timehist_th}a shows the r.m.s.~temperature fluctuations close to the ground at $z^+\approx15$.
The large peaks around $tU_\star/h\approx10$ for the dry cases 395D06 and 590D085 (solid red and green lines, Fig.~\ref{fig:timehist_th}a) are due to spatial intermittency of turbulent/laminar patches, which are locked in place due to the periodic boundary conditions \citep{Flores11,GarciaVillalba11}.

Figure \ref{fig:timehist_th}b shows the  temperature fluctuations in the outer layer at $z/h=2/3$.
As discussed with Fig.~\ref{fig:timehist_flux_beta}, the downwards heat flux has not reached a statistically steady state
and so we see $T'_{rms}$ increases with time for both the dry and moist cases.
Critically, by normalizing on the friction temperature $T_\tau$ defined with 
varying Bowen ratio $\beta$, we see that the temperature fluctuations in the
moist case  do not vary when the saturation interface reaches this level. 
The fluctuations remain similar to the dry cases,
despite the moist case having both intermittent  unsaturated and saturated regions 
as the saturation interface reaches this level (which occurs around $tU_\star/h\approx10$--20 for cases 395M041 and 395M06). Moreover,
once the air at $z/h=2/3$ becomes completely saturated
after approximately 20 large-eddy turnover times
the moist temperature fluctuations still remain similar to the dry cases.
This suggests that these fluctuations within the saturated air are similar
to those in the dry air when appropriately normalized.

\subsection{Fog Development}
\label{ssect:fogdev}

\setlength{\unitlength}{1cm}
\begin{figure}
\centering
  \includegraphics{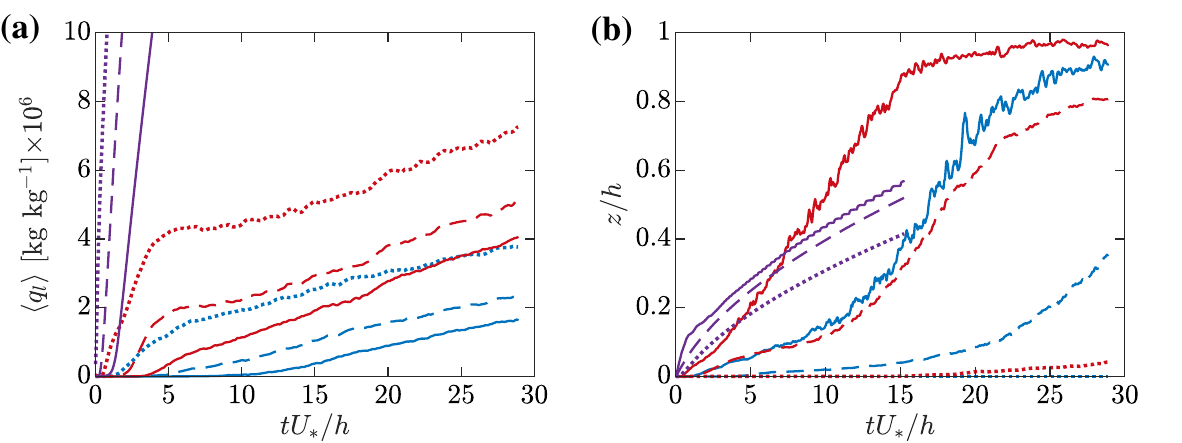}
  \vspace{-0.05cm}
\caption{
(a) Time series of the  liquid water mixing ratio at different heights:
dotted, $z^+\approx 0$;
dashed, $z^+\approx 15$;
solid, $z/h\approx 0.1$.
(b) Time series of the height where $\langle q_l\rangle$ is equal to a specified threshold, $q_{l,t}$:
dotted, $q_{l,t}=5\times 10^{-6}$;
dashed, $q_{l,t}=1\times 10^{-6}$;
solid, $q_{l,t}=10^{-8}$ (saturation interface) kg kg$^{-1}$.
Line colours denote cases:
blue, 395M041 (weakly stable);
red, 395M06 (weakly stable);
purple, 395M205 (very stable)
}
\label{fig:qcprofs}
\end{figure}

Before analyzing the liquid water mixing ratio, $q_l$, we note that values of $q_l$ can be converted to visibility values according to the empirical relationship from \cite{Kunkel84},
\begin{equation}
VIS(z,t)=-\f{\log(0.02)}{144.7(\rho \langle q_l\rangle)^{0.88}},
\end{equation}
 where $VIS$ is measured in metres.
The visibility values are typically much larger than the surface layer height and therefore domain size ($VIS \gg h$),
and this empirical relationship is unlikely to be entirely appropriate for the present idealized system.
As such, we only use $VIS$ for providing simple, qualitative descriptions in the context of fog.
For reference, values of $q_l=1\times 10^{-6}$ and $5\times 10^{-6}$  kg kg$^{-1}$ correspond to visibilities of approximately 4 km and 1 km, respectively.
These two values will be subsequently used  to track the height where  the horizontally averaged liquid water mixing ratio is equal to these threshold values.
 Recall that fog is classified as such when the visibility falls below 1 km \citep{NOAA17}.

A time series of the liquid water mixing ratio at different heights is shown in Fig.~\ref{fig:qcprofs}a,
for both weakly stable cases (395M041 and 395M06) as well as a very stable case (395M205).
For the weakly stable cases
the flow nears laminarization in the first 5 eddy turnover times,
which leads to the liquid water mixing ratio increasing rapidly.
This is most obvious close to the ground for the case with the larger cooling rate, 395M06 (black dashed line, Fig.~\ref{fig:qcprofs}a).
Thereafter, $\langle q_l\rangle$ increases linearly with time for both cases and at all heights, although due to the larger cooling rate for case 395M06 the rate of increase of $\langle q_l\rangle$  is larger than for case 395M041 (solid lines).
In the very stable case (dotted lines), the reduced mixing of the laminar flow causes $\langle q_l\rangle$ to increase significantly over a very short space of time, as well as at heights further from the ground. The liquid water mixing ratio reaches values of $60\times 10^{-6}$ and $40\times 10^{-6}$ kg kg$^{-1}$ at $z/h\approx0$ and $z/h\approx 0.1$, respectively, at time $tU_\star/h\approx15$, although this is not shown for clarity.

Figure \ref{fig:qcprofs}b tracks the height where the horizontally averaged liquid water mixing ratio, $\langle q_l \rangle$ is equal to a specified threshold, where we take the threshold values of $q_{l,t}=1\times 10^{-6}$ and $5\times 10^{-6}$ kg kg$^{-1}$ mentioned above.
Also shown is the interface between saturated and unsaturated air, obtained using a threshold of $q_{l,t}=10^{-8}$ kg kg$^{-1}$.
Note that choosing a different saturation interface threshold of, for example, $q_{l,t}=10^{-9}$ kg kg$^{-1}$ does not significantly alter the results presented in Fig.~\ref{fig:qcprofs}b as these are essentially all measures of the point where $\langle q_v\rangle=q_{sat}$.
For the weakly stable cases the saturation interface rises rapidly from the ground and 
for case 395M06 it reaches the top of the domain ($z/h=1$)
at around $tU_\star/h\approx15$. 
Due to the impermeability constraint with zero moisture flux, this top boundary would therefore
now influence the development of liquid water. Indeed, we see the height based on $q_{l,t}=1\times10^{-6}$ kg kg$^{-1}$ ($VIS\approx 4$ km, dark grey dashed line) increases rapidly at this point,
as the turbulent mixing is  now bringing already saturated air from aloft, rather than mixing unsaturated and saturated air as was previously the case.
The flow after $tU_\star/h\approx15$ for this case is therefore unlikely to be physical due to the influence of the top boundary. 
Nevertheless, at $tU_\star/h\approx15$  the liquid water mixing ratio immediately adjacent to the ground is approximately $5\times10^{-6}$ kg kg$^{-1}$, corresponding to $VIS\approx 1$ km and  could thus be classified as fog. 
If a similar simulation were conducted of a turbulent Ekman layer then this would not be subject to the top boundary influence at $tU_\star/h\approx15$. We would therefore expect the liquid water mixing ratio to continue to increase and as such show an increasingly larger region of fog.

The very stable case, 395M205, shows that the visibility reduces  significantly close to the ground.
The saturation interface (light grey dotted line) grows at a similar rate compared to the weakly stable cases.
Meanwhile, the height where $\langle q_l\rangle$ is equal to either $q_{l,t}=1\times10^{-6}$ or $5\times10^{-6}$ kg kg$^{-1}$ grows much more rapidly than the weakly stable cases. The latter threshold, along with Fig.~\ref{fig:qcprofs}a, suggests that  large regions of the domain are filled with sheets of fog.
This is due to the runaway cooling effect \citep{vandeWiel07}, wherein  the lack of turbulent mixing causes
the temperature to rapidly reduce and thus substantial condensation to occur.
Here, the imposed ground cooling is approximately 3.4 times greater than the weakly stable case 395M06, yet the values of $\langle q_l\rangle$ are 
over an order of magnitude larger.

\setlength{\unitlength}{1cm}
\begin{figure}
\centering
  \includegraphics{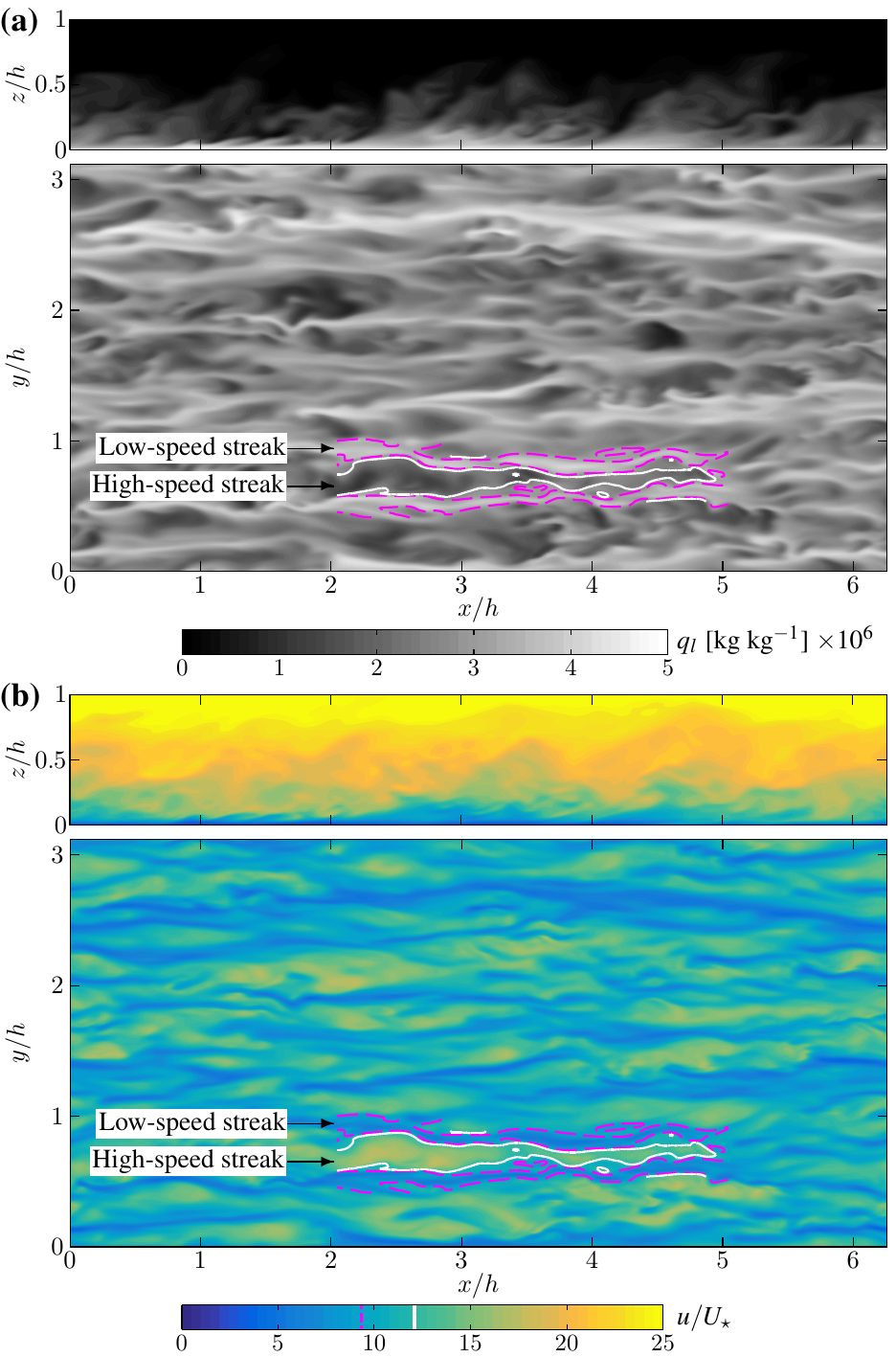}
    \vspace{-0.05cm}
\caption{Instantaneous visualization of 
(a) liquid water mixing ratio 
and 
(b) streamwise velocity, 
in the streamwise--vertical ($x$--$z$) plane
 and horizontal ($x$--$y$)  plane at $z^+\approx 15$,
for case 395M06 at $tU_\star/h\approx 15$. 
Solid white and dashed magenta contours in lower-centre region of the horizontal planes (between $2\leq x/h\leq5$) show $\langle u\rangle\pm0.5u'_{rms}$ to highlight a few high- and low-speed streaks.
Mean flow is from left to right
}
\label{fig:viz}
\end{figure}

\begin{figure}
\centering
 \includegraphics{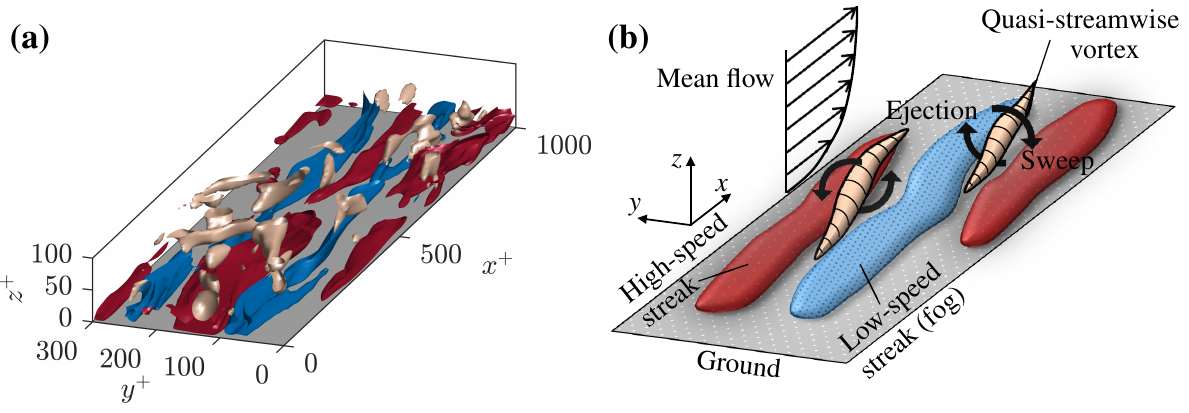}\\
  \vspace{-0.05cm}
\caption{
(a) Snapshot showing isosurfaces of 
high- and low-speed streaks ($u^+-\langle u^+\rangle = \pm3$, red and blue respectively) 
and quasi-streamwise vortices ($|w^+| = 2$, tan). 
Structures above $z^+=60$ are removed.
(b) Sketch of the near-ground low- and high-speed streaks accompanied by quasi-streamwise vortices.
Increased liquid water content (fog) concentrates in the low-speed streaks
}
\label{fig:streakfogsketch}
\end{figure}

Figure \ref{fig:viz} shows an instantaneous snapshot of the liquid water mixing ratio and streamwise velocity at $tU_\star/h\approx15$ for the turbulent case 395M06. This is the point
at which almost the entirety of the air is at least somewhat saturated, indicating that further development of liquid water is unphysical due to the subsequent influence of the
top boundary. 
The vertical cross-section of liquid water shows coherent structures that are inclined toward the horizontal, similar to the `ramp-like' velocity structures observed in wall-bounded turbulence
\citep{Adrian00,Marusic07}.
The horizontal cross-section in a plane close to the ground at $z^+\approx15$ shows
 that the liquid water and streamwise velocity contours are well correlated,
where
the highest liquid water content (or lowest visibilities) are associated with the streamwise-elongated streaks of low-speed streamwise velocity
and conversely the low liquid water regions (higher visibility) are associated with high-speed streaks.
These streaks \citep{Kline67} are accompanied by shorter quasi-streamwise vortices \citep{Jeong97}.
A snapshot showing isosurfaces of the  streaks ($u^+-\langle u^+\rangle = \pm3$)  and quasi-streamwise vortices ($|w^+| = 2$) is given in Fig.~\ref{fig:streakfogsketch}a,
where these isotach thresholds  are taken from \cite{Jimenez18}.

The streaks and quasi-streamwise vortices form the basis of the near-wall cycle,
which is the well-documented self-sustaining mechanism that generates turbulence in the buffer region \citep{Hamilton95,Jimenez99,Schoppa02,Jimenez18}.
A sketch is presented in Fig.~\ref{fig:streakfogsketch}b showing how this mechanism relates to fog.
The quasi-streamwise vortices draw cooler air with reduced momentum away from ground in a process called an ejection. These 
form low-speed streaks and contain more liquid water.
Meanwhile, the vortices also cause a downwash of warmer, drier air from aloft towards the ground. This process, termed a sweep, forms the high-speed streaks.
The streaks  meander perpendicular to the flow direction and eventually break down in a process called bursting,
leading to vorticity generation  and regeneration of the quasi-streamwise vortices.
This therefore completes the self-sustaining cycle \citep{Hwang16,Jimenez18}.

In the context of fog, the ejections enable the saturation interface (Fig.~\ref{fig:qcprofs}b) to grow quickly with time,
as this mechanism rapidly draws  liquid water away from the ground.
However, the sweeps draw down drier air so that the horizontally averaged liquid water mixing ratio  does not increase as rapidly as in the laminar case.
This view is consistent with studies of particle-laden turbulent flows, 
where high concentrations of particles were observed to accumulate within low-speed streaks \citep[e.g.][]{Rashidi90,Pan96,Soldati09,Lee15}.
Note that the higher Reynolds number cases with $Re_\star=590$ include the same mechanism to that displayed in Fig.~\ref{fig:streakfogsketch} (not shown).
Furthermore, in neutrally stratified flows a similar set of self-sustaining streaky structures appear higher up in the boundary layer, 
albeit at larger length scales on the order of $h$
and with bursting time scales of $tu_\tau/h\sim6$ 
\citep{Flores10,Hwang15,Cossu17, MacDonald17}.
This suggests that the mechanism for fog formation shown in  Fig.~\ref{fig:streakfogsketch} is not limited to the near-ground region or to the low Reynolds numbers of the present simulations.
We finally note that these streaky structures are somewhat reminiscent of the submeso motions that have been identified in SBL observational studies \citep{Mahrt14}.
Under conditions of very weak shear with $u_\tau\approx 0.02$ m s$^{-1}$
and a surface layer height of $h\approx 1$ m \citep{Flores11}
we would expect the bursting period to be of the order of several minutes,
in agreement with the time scales associated with submeso (or at least hybrid) motions \citep{Mahrt14}.

\setlength{\unitlength}{1cm}

\subsection{Turbulence Collapse in Dry and Moist Flows}
\label{ssect:turbcollapse}

\setlength{\unitlength}{1cm}
\begin{figure}
\centering
  \includegraphics{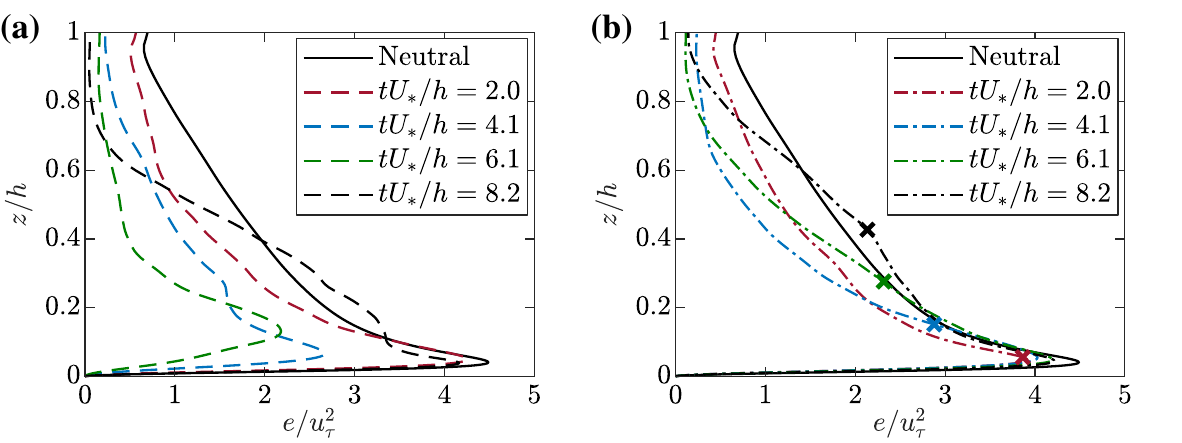}\\
   \vspace{-0.05cm} 
\caption{Profiles of 
TKE
at different times during the initial fog formation stage,
for 
(a) 395D06, 
and 
(b) 395M06.
The solid black line represents the statistically steady dry neutrally stratified case.
The $\boldsymbol{\pmb{\times}}$ symbols in (b)  indicate the height of the saturation interface (Fig.~\ref{fig:qcprofs}b)
}
\label{fig:tkeprofs}
\end{figure}


As discussed above, the weakly stable cases approach a laminar state before returning to be completely turbulent.
In order to investigate this process, Fig.~\ref{fig:tkeprofs} shows the profiles of TKE
at different
times when liquid water first condenses. 
This is non-dimensionalized on the instantaneous friction velocity, $u_\tau$, at the corresponding times,
as \cite{Flores18} shows the flow scales with instantaneous $u_\tau$ close to the ground for weak stratification.
Initially, $e$ is close to the neutral case at $tU_\star/h=2.0$ for both dry and moist cases. Thereafter, for the dry case,
the TKE significantly reduces,  with the peak value close to half of that of the neutral case.
At $tU_\star/h=8.2$, the dry TKE has returned to a similar value of the neutral case, although there is very
little energy above $z/h\approx0.8$.
Conversely, in the moist case (Fig.~\ref{fig:tkeprofs}b), the TKE remains similar to
 the neutral case. There is some variation in the outer layer, although the peak value close to $z/h\approx0.04\Rightarrow z^+\approx15$ 
 consistently remains above $e/u_\tau^2\approx 4$. 
 Over this time period, the saturation interface grows
 from $z/h\approx0.05$ at $tU_\star/h=2$ to $z/h\approx 0.41$ at $tU_\star/h= 8.2$ (shown by the $\boldsymbol{\pmb{\times}}$ symbols).
 The condensation of liquid water therefore appears to produce significant TKE, agreeing with the `burst' of TKE observed in  LES studies at the very onset of fog formation \citep{Nakanishi00, Bergot13}.

\setlength{\unitlength}{1cm}
\begin{figure*}
\centering
  \includegraphics{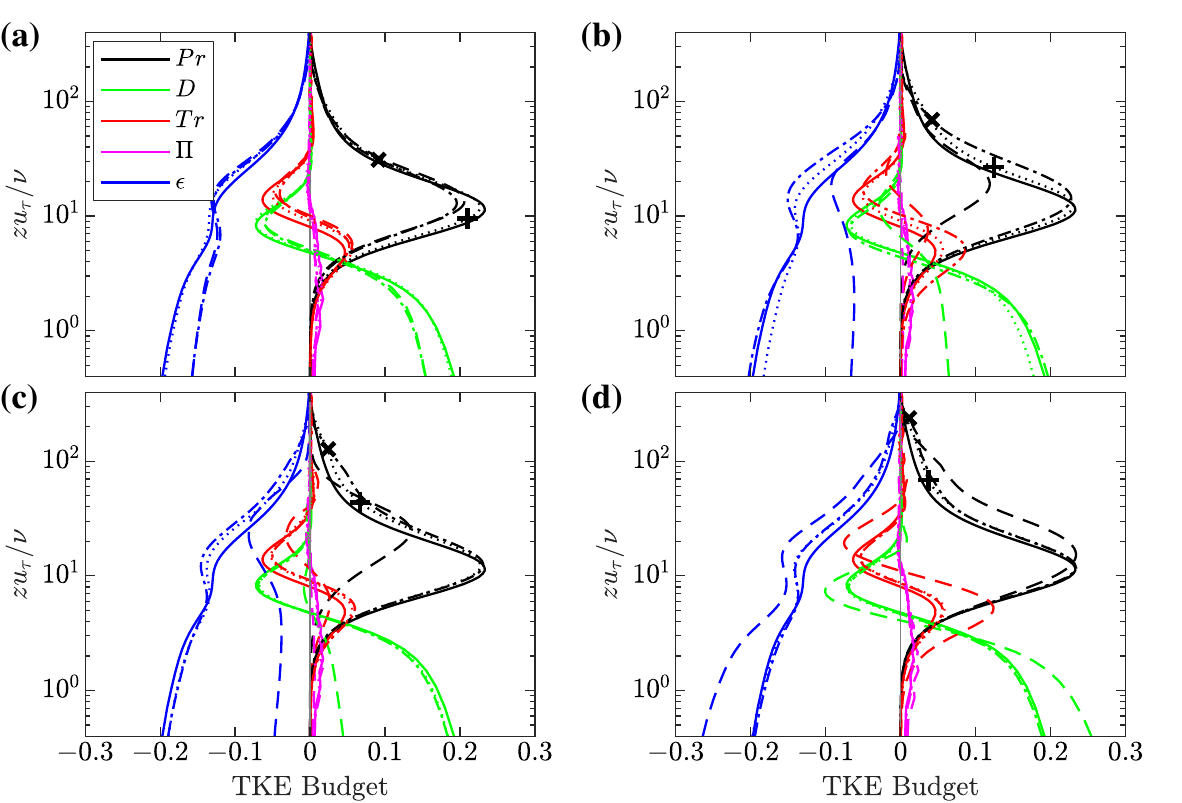}
  \vspace{-0.05cm}
\caption{
TKE budget 
for
395D00 (solid),
395M041 (dotted),
395D06 (dashed)
and
395M06 (dash-dotted),
 at  times (a)  $tU_\star/h= 2.0$, (b)  $tU_\star/h= 4.1$, (c)  $tU_\star/h= 6.1$  and (d) $tU_\star/h= 8.2$, as in Fig.~\ref{fig:tkeprofs}.
The budget terms are defined in 
Eq.~\ref{eqn:tkebudget} and are normalized on the instantaneous friction velocity, $u_\tau$, and viscosity, $\nu$.
The black $\boldsymbol{\pmb{+}}$ and $\boldsymbol{\pmb{\times}}$ symbols indicate the height of the saturation interface 
(Fig.~\ref{fig:qcprofs}b) 
for 395M041 and 395M06, respectively.
The buoyant production term, $B$, is small and not shown
}
\label{fig:tkebudg}
\end{figure*}

The TKE budget is shown in Fig.~\ref{fig:tkebudg} for the same times as Fig.~\ref{fig:tkeprofs},
where the budget terms are normalized on the instantaneous friction velocity, $u_\tau$ and viscosity $\nu$.
Note that due to the transient nature of these cooling cases and variations in TKE, the tendency or residual term is non-zero;
however,  it is typically of similar magnitude to the pressure correlation term, and thus is not shown for clarity.
Similarly, the buoyant production of TKE, $B$, is small and not shown.
The small magnitude of $B$ has been discussed previously in DNS studies of stably stratified uniformly sheared flow \citep{Jacobitz97} and  Ekman layers \citep{Shah14,Ansorge14}. There it has been suggested that the main influence of stable stratification on TKE is in reducing the vertical velocity variance, rather than directly through buoyant destruction of TKE. 

Initially, at $tU_\star/h=2.0$ (Fig.~\ref{fig:tkebudg}a), both the dry and moist TKE budget terms for $h/L=0.6$ (dashed and dash-dotted lines, respectively) are almost identical and in moderate agreement with
the neutral case.
However, at times $tU_\star/h=4.1$ and 6.1 (Fig.~\ref{fig:tkebudg}b, c), the dry TKE production has significantly reduced compared to
the neutral and moist cases. 
This reduction is somewhat balanced by a reduction in dissipation, as was also noted for the horizontal and vertical kinetic energies in \cite{Flores18}.
The formation of liquid water in the moist case, where the height of the saturation interface  is shown by the $\boldsymbol{\pmb{\times}}$ symbols,
appears to maintain the production of TKE similar to that of the neutral case. However, at $z^+\approx15$ the dissipation of the moist case is always greater in magnitude
when compared to the neutral case for all times. Even at $tU_\star/h=8.2$, when the dry and moist cases are returning to similar balances as the neutral case,
the dissipation always remains somewhat larger than in the neutral case at this vertical location. 

The moist case with a smaller cooling rate, $h/L=0.41$ (dotted lines, Fig.~\ref{fig:tkebudg}), is observed to maintain an instantaneous equilibrium and remain
similar to the neutral case throughout, although there is enhanced dissipation close to $zu_\tau/\nu\approx15$. There is also slightly larger
production in the outer layer for this moderate cooling case, as observed by \cite{Nieuwstadt05} and \cite{Flores18}.
 The dry case for $h/L=0.41$ is similar to the moist case and so is not shown in Fig.~\ref{fig:tkebudg}.
 The condensation of liquid water therefore only appears to have a significant impact during
the transient adjustment to the cooling rate, when the flow is close to laminarization.

\setlength{\unitlength}{1cm}
\begin{figure}
\centering
  \includegraphics{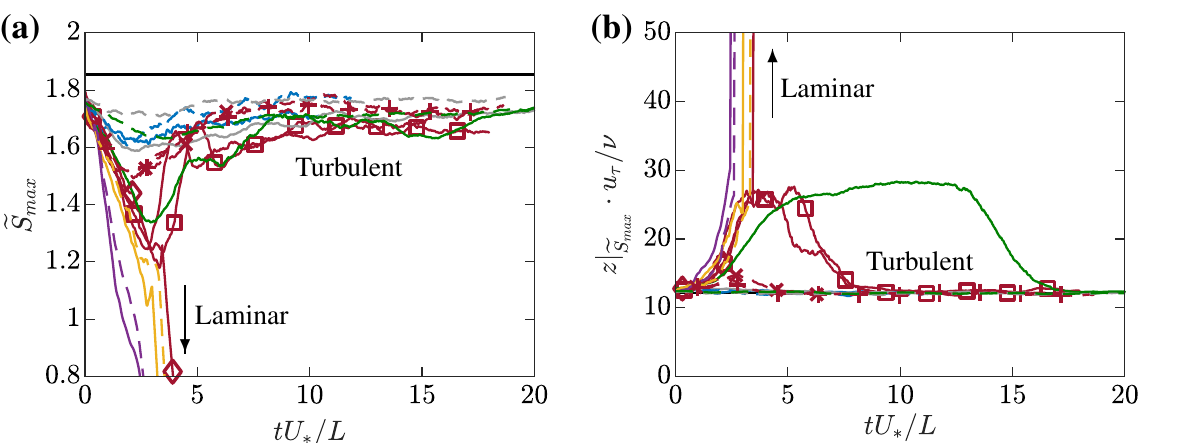}\\
  \vspace{-0.05cm}
\caption{Time series of (a) the maximum non-dimensional shear rate $\widetilde{S}_{max} = \max((Pr+B)/\epsilon)$
and
(b) the vertical position of this maximum, non-dimensionalized on the instantaneous friction velocity $u_\tau$.
Line styles same as Fig.~\ref{fig:timehist_wd}.
The mean steady state values of dry neutral case 395D00 are given by the horizontal black lines.
Data are only shown when $Pr\nu/u_\tau^4>0.01$ (i.e., the flow is turbulent)
}
\label{fig:streakstrength}
\end{figure}

Finally, we investigate the effect of condensation of liquid water on the high- and low-speed streaks observed in Fig.~\ref{fig:viz}.
As mentioned earlier, these streaks are a fundamental mechanism in generating turbulence in the buffer layer.
The ratio between the instantaneous, horizontally averaged production and dissipation of  TKE, $\widetilde{S}(z,t)=(Pr+B)/\epsilon$, or non-dimensional local shear rate,
was found to be related to the existence of streaks in \cite{Lam92}.
When $\widetilde{S}<1$ and dissipation exceeds production, the low-speed streaks were no longer observed.
Conversely, when production increases and $\widetilde{S}\gtrsim1$ the streaks were found to be increasingly energetic and persistent with increasing $\widetilde{S}$.

Figure \ref{fig:streakstrength}a shows the evolution of the maximum of $\widetilde{S}$,
where data are only shown when the maximum production $Pr\nu/u_\tau^4>0.01$, or 5\% of the maximum production in the dry neutrally stratified case.
This is done because $Pr$ and $\epsilon$ tend to zero in the cases where turbulence collapses, making $\widetilde{S}$ undefined.
Before turbulence collapses, in the laminar cases we see that $\widetilde{S}$ tends to below the critical value $\widetilde{S}\approx 1$.
The streaks therefore cannot remain self-sustaining and the near-wall cycle is destroyed.
In the cases that remain turbulent, $\widetilde{S}$ initially reduces  as the flow adjusts to the imposed cooling,
before tending towards a steady state value that is nearly 10\% less than the mean of the dry neutrally stratified case.
This is due to enhanced dissipation in the cooling cases, as observed from the TKE budgets (Fig.~\ref{fig:tkebudg}). 
For moderate cooling rates of $h/L=0.41$, little difference is seen between the dry and moist cases.
However, for the weakly stable cases with stronger cooling of $h/L=0.6$ for both $Re_\star=395$ and 590, $\widetilde{S}$ is slightly less for the dry case than for the moist case.
This suggests that the condensation of liquid water energizes the near-wall cycle, promoting streaks and thus offering a possible explanation for how the flow is able to sustain turbulent
motions at higher cooling rates than the dry case. From the TKE budgets in Fig.~\ref{fig:tkebudg}, this appears to be  due 
to
reduced dissipation in the moist case compared to the dry case,
rather than increased TKE production.
 Neutrally stratified turbulent flows laden with heavy particles have also been seen to energize turbulent flows, although only for low Stokes number particles \citep{Lee15}. There, the mechanism
is slightly different as the inertia of the particles imparts momentum on the fluid and increases TKE production, rather than reduced dissipation observed in the moist cases in this study.

Figure \ref{fig:streakstrength}b shows the vertical position of the maximum local shear rate, $z|_{\widetilde{s}_{max}}$, non-dimensionalized on the instantaneous friction velocity $u_\tau$. 
This indicates the position where streak formation is dominant; for the present dry neutral case this is located at approximately $z^+\approx12$, in agreement with \cite{Lam92}.
The weakly stable cases also show that the streak formation remains located at approximately $zu_\tau/\nu\approx 12$.
For the very stable cases, the vertical position where streak formation is strongest is progressively pushed away from the ground, presumably until
the reduced shear higher up cannot sustain streak generation and the turbulence collapses.

\section{Conclusions}
\label{sect:conc}
The effects of moisture on the stably stratified surface layer are investigated using direct numerical simulations (DNS),
with particular attention given to how this system can be treated as an idealized representation of  fog formation.
A cooling heat flux is imposed at the ground yielding both weakly stable (turbulent)  and very stable (laminar) solutions,
when initialized from a neutrally stratified turbulent surface layer.
The relative humidity is initialized close to 100\%, so that as the air cools it reaches its saturation point and liquid water condenses.
 In the weakly stable moist cases,
the vertical height of the saturation interface grows quickly until it reaches
the top of the domain.
However, due to turbulent mixing the liquid water mixing ratio remains relatively small throughout the domain, with only the lowermost 
few percent of the surface layer achieving enough liquid water to correspond to visibilities close to 1 km.
In the very stable cases the liquid water mixing ratio increases rapidly.
While the saturation interface grows at a similar rate to the turbulent cases,
the visibility significantly reduces below 1 km  due to runaway cooling.
This idealized system suggests that while turbulence and its associated mixing does not inhibit fog formation, it does impede its growth.

In the case where the ground cooling is close to its critical value ($h/L\approx 0.6$ for $Re_\star=hU_\star/\nu=395$),
two dry cases with unique initialization snapshots were found to sustain turbulent motions, while in another
case with a different initialization snapshot it did not. 
Meanwhile, when moisture effects were enabled, the flow was  able to sustain turbulent motions when initialized from the same three snapshots.
Thus, for
the moist case the critical cooling value is between $0.6<h/L<0.7$
while for the dry case it was at most $h/L\lesssim0.6$
 for $Re_\star=395$.
The moist cases were not as close to laminarization, with the inner-normalized local Obukhov length $L_\tau u_\tau/\nu$ always exceeding a value 225, while turbulence collapsed in the dry cases when $L_\tau u_\tau\nu\lesssim 100$ \citep{Flores11}.
 The  latent heat release through condensation appears to enhance
TKE relative to the dry case, enabling the moist cases to sustain slightly larger ground cooling fluxes
(or higher Richardson numbers) before leading to turbulence collapse. 
We hypothesize that any tendency towards laminarization with reduced mixing would result in enhanced cooling and thus lead to condensation, producing 
additional mixing that would avert,  or at least delay, complete turbulent collapse.

Visualization of the weakly stable moist cases reveal that regions of increased liquid water mixing ratio develop within the low-speed streaks
associated with the well-documented near-wall cycle. These low-shear regions
draw slower-moving air up from closer to the ground, 
and suggest a mechanism for how fog can form within turbulent flow.
In neutrally stratified flows these streaky structures have also been  observed at higher Reynolds number as well as at larger scales higher up in the boundary layer \citep{Flores10,Hwang15,Cossu17}, and are  somewhat reminiscent of the submeso motions in SBL observations \citep{Mahrt14}.
This suggests that the mechanism for fog formation discussed in this paper, obtained from simulations of an idealized system at relatively modest Reynolds numbers, should also apply to the real atmosphere.

While the streaks of both dry and moist weakly stable cases are slightly weaker than in the neutrally stratified case,
the moist case appears to produce streaks which are slightly more persistent and energetic than in the dry case.
This is due to the dissipation of TKE being greater in the dry case 
than the moist case, although both are larger than in the neutral case. The shear production of TKE for the moist case
appears to remain somewhat similar to the neutrally stratified case, 
although when saturation first occurs it results in a slight increase in production relative to the neutral case.


\begin{acknowledgements}
This research was carried out at the Jet Propulsion Laboratory, California
Institute of Technology, under a contract with the National Aeronautics and Space
Administration.
Parts of this research were supported by the U.S. Department of Energy, Office of Biological and Environmental Research, Earth System Modeling; the NASA MAP Program; the Office of Naval Research, Marine Meteorology Program and the NOAA/CPO MAPP Program.
We thank the three anonymous reviewers for their constructive comments on the manuscript.
The authors also acknowledge the Texas Advanced Computing Center (TACC) at The University of Texas at Austin for providing HPC resources that have contributed to the research results reported within this paper.

\end{acknowledgements}


\section*{Appendix 1: Effect of Computational Domain Size}
In the present study, the domain size in the streamwise and spanwise directions is  $2\pi h\times \pi h$, 
which is standard for neutrally stratified dry simulations \citep{Lozano14,Munters16}.
However, 
there can exist patches of laminar and turbulent flow  in stably stratified flows close to laminarization, which are of  size $10h$ (that is, the present domain size). 
A small computational domain can therefore result in turbulent structures that get `locked' in place due to the periodic boundary conditions \citep{Flores11,GarciaVillalba11}.
This can require substantial time for the turbulence to repopulate the domain and does not represent a physically realistic scenario.

\begin{figure}
\centering
  \includegraphics{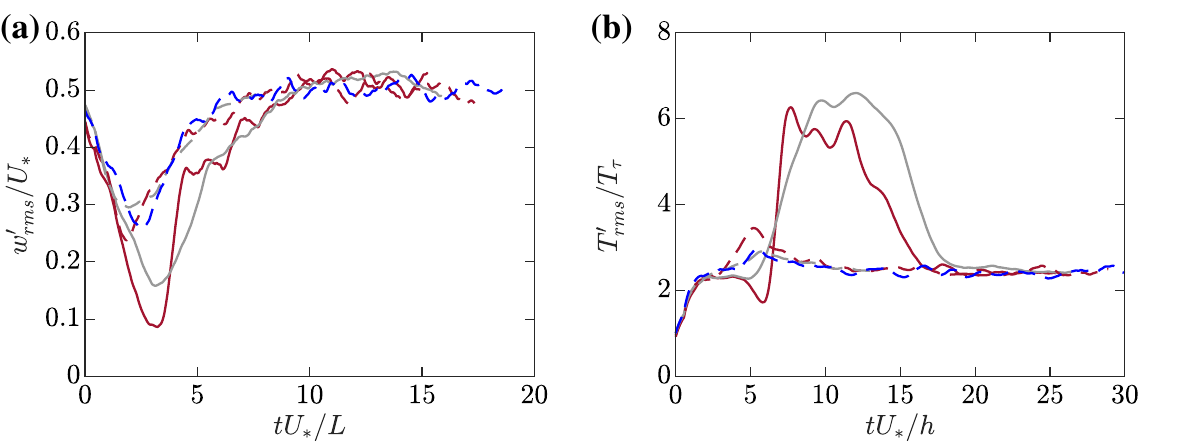}\\
  \vspace{-0.05cm}
\caption{
Time series of 
(a) r.m.s.~vertical velocity  fluctuations
and
(b) temperature fluctuations
at $z^+=15$.
Line styles are:
red, $h/L=0.6$ with $T_0=279$ K (base case);
grey, base case except with enlarged domain;
blue, base case except with $T_0=285$ K.
Solid lines denote dry cases, dashed denote moist cases
}
\label{fig:timehist_app}
\end{figure}

We double the domain size to $4\pi h\times2\pi h$ in the streamwise and spanwise directions, with the results shown in 
Fig.~\ref{fig:timehist_app} 
for cooling with $h/L=0.6$  and $Re_\star=395$. 
We see a similar behaviour for both dry (solid) and moist (dashed) cases compared to the regular domain size.
 The peak around $tU_\star/h\approx10$ for the dry case temperature fluctuations 
 (Fig.~\ref{fig:timehist_app}b) 
 is due to the turbulence becoming locked in place.
 Visual inspection of the 
velocity field of the dry case (not shown) reveals a similar pattern to Fig.~5 of \cite{Flores11}, in which there exist stripes of turbulent and laminar flow extending down the entire streamwise length of the domain.
The present larger domain dry case also exhibits this effect, where \cite{GarciaVillalba11} notes domains of at least $8\pi h\times 3\pi h$ are required to observe intermittency in the streamwise direction for statistically steady flows.
This behaviour is not observed in any of the moist case, as presumably the mixing induced by condensation is sufficient to completely avoid these large laminar patches at these cooling rates.
In this work, as in \cite{Flores11}, we ignore data from when the turbulent flow is locked in place, as it is artificial. This is primarily of concern for the dry cases, and was not observed for the moist cases.

\setlength{\unitlength}{1cm}

\section*{Appendix 2: Effect of Initial Temperature}

\setlength{\unitlength}{1cm}
\begin{figure}
\centering
  \includegraphics{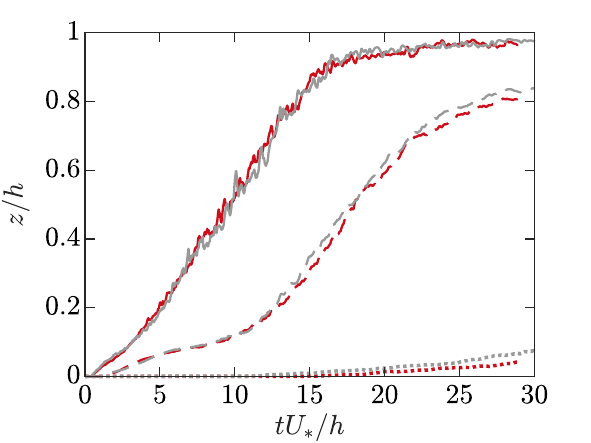}
  \vspace{-0.05cm}
\caption{
Same as Fig.~\ref{fig:qcprofs}b, showing
time series of  
the height where $\langle q_l\rangle$ is equal to a specified threshold, $q_{l,t}$:
dotted, $q_{l,t}=5\times 10^{-6}$.
dashed, $q_{l,t}=1\times 10^{-6}$;
and
solid, $q_{l,t}=10^{-8}$ (saturation interface) kg kg$^{-1}$,
for  the weakly stable case 395M06 with $h/L=0.6$.
Line colours:
red, $T_0=279$ K (base case);
grey, $T_0 = 285$ K
}
\label{fig:qcprofs_app}
\end{figure}

Given that the liquid water mixing ratio depends non-linearly on temperature, we also investigate the effect of the temperature at which the simulations are initialized, $T_0$.
For the cases studied in this paper, we set $T_0=279.15$ K, while here we increase the initial temperature to 285.15 K, for matched cooling rates of $h/L=0.6$. 
Figure \ref{fig:qcprofs_app} 
shows the time series of fog height, as in Fig.~\ref{fig:qcprofs}b for these two initial temperatures.
The two cases are initially similar although they start to diverge after approximately $tU_\star/h\gtrsim15$,  at which point the interface between
saturated and unsaturated water has reached the top boundary.
The vertical velocity and temperature fluctuations shown in 
Fig.~\ref{fig:timehist_app} 
also show good agreement between the base case (red dashed line) and increased initial temperature case (blue dashed line).
This suggests that, while the temperature will be important for specific meteorological events, it is likely not relevant here given the idealized system studied in this paper.

%


%
%
%
%

\bibliographystyle{spbasic_updated}      
\bibliography{bibliography}   

\end{document}